# Phase Transitions in Lattice QED

M. Baig[a]

[a]Grup de Física Teòrica, IFAE
Universitat Autònoma de Barcelona
08193 Bellaterra (Barcelona) SPAIN

The main objective of the work presented here is to understand the appearance of phase transitions in pure gauge and scalar lattice QED, both in the compact as well as the noncompact formulations.

## 1. INTRODUCTION

Recent developments in the study of lattice QED without fermions have lead to a rather puzzling scenario

- Pure gauge compact QED with periodic boundary conditions (PBC) shows a first order phase transition associated to a monopole condensation phenomena [1,2].

- Pure gauge non compact QED with PBC is a gaussian model that does not exhibits phase transitions, but shows monopole percolation [3].

- Non-compact QED with PBC coupled to unitary Higgs fields shows two different lines, one of second order phase transitions and another related to a monopole percolation phenomena [4].

We will present here four topics that are intended to clarify some of the relations between monopoles and phase transitions.

## 2. MONOPOLE PERCOLATION

The lattice action used in our simulation [5] is the original Wilson (compact) action

$$S_{gauge} = \sum_{n\mu\nu} \cos\Theta_{\mu\nu}(n). \qquad (1)$$

We have applied the standard Metropolis algorithm to simulate the compact U(1) lattice gauge theory with standard periodic boundary conditions on lattices from $8^4$ to $14^4$. We have made measurements of the monopole observables near the phase transition.

The evolution of the internal energy, the cluster susceptibility $\chi$ and the $n_{max}/n_{tot}$ order parameter in "Monte Carlo time", i.e. the iteration number, show a clear correlation. The "tunneling" between the two phases is very clear for all three observables at the same point. This phenomena is characteristic of a first order phase transition.

Our conclusion is that in compact pure gauge QED, the percolation threshold occurs at the same point as the deconfining phase transition, i.e. the point of monopole condensation. Moreover, the $n_{max}/n_{tot}$ order parameter used to characterize monopole percolation proves to be an excellent order parameter to study the location of the apparent first order phase transition.

## 3. FIXED BOUNDARY CONDITIONS

Numerical simulations of lattice gauge theories are actually performed in a finite space-time box, and computer limitations imply severe restrictions on the lattice size of this box. To avoid the border "effects" generated by such a box, periodic boundary conditions are usually adopted. From a geometrical point of view, this fact implies that the system is actually simulated on an hypertorus.

A possible explanation for the observed discontinuity in the internal energy on pure gauge compact QED with PBC relies on the fact that topologically nontrivial loops which wrap around the lattice are permitted. Remember that in $D = 3$ monopoles are 0-dimensional pointlike excitations whilst in $D = 4$ they become 1-dimensional, and, due to the magnetic flux conservation, they form closed loops.

In order to avoid toroidal topology we have imposed to our lattice fixed boundary conditions (FBC). This means that we fix -and keep- to the unity all links belonging to the boundary of the lattice. From a geometrical point of view this is equivalent to a collapse to a single point all the boundary obtaining an effective spherical topology.

Our numerical results [6] come from some thermal cycles performed for lattice sizes $6^4, 8^4, 10^4, 12^4, 14^4$ and $16^4$. No sign of discontinuity has been observed. In addition, the emergence of a peak of the specific heat which grows with $L$ and moves towards $\beta = 1$ has been clearly observed. Also the growing of that peak seems slow enough to ensure the absence of a weak first order transition. Furthermore, the Binder cumulant shows a clear convergence to the expected value of 2/3 characteristic of a second order transition.

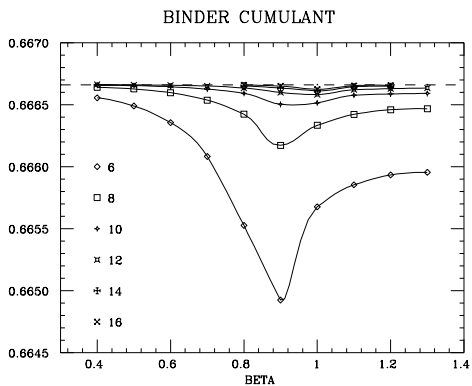

Figure 1. Binder cumulant measurement.

These results, although being basically qualitative, point out the continuous character of the U(1) phase transition when periodic boundary conditions and the spurious effects they bear with are avoided.

## 4. NON-COMPACT SCALAR QED

The non-compact coupled gauge-Higgs model (scalar QED) has been extensively studied in a recent paper [4]. Our model has gauge fields in the links of the lattice that are elements of the $U(1)$ gauge group $\theta_{x,\mu}$ and matter fields on the sites $\phi_x = e^{i\alpha(x)}$. The action is

$$S = \frac{1}{2}\beta \sum_p \theta_p^2 - \lambda \sum_{x,\mu}(\phi_x^* U_{x,\mu} \phi_{x+\mu} + cc), \qquad (2)$$

where $\theta_p$ is the circulation of the gauge field around a plaquette, $\beta = \frac{1}{e}$ is the gauge coupling and $\lambda$ the Higgs coupling. Phase diagram shows two separate regions, a Coulomb phase and a Higgs phase separated by a second order transition line. In the limit $\gamma \to \infty$ limit the transition, however, becomes of first order. This limit can be interpreted as an Integer Gaussian Model (IGM) since the action becomes

$$S_\infty = -\frac{\beta}{2} \sum_p (2\pi n_p)^2. \qquad (3)$$

It is interesting to note that eq. 3 is just the Villain action for the pure gauge compact QED but with a $2\pi^2 \beta$ instead of $1/\beta_v$. Moreover, if one consider the Lagrangian version of the Hamiltonian Loop Representation for the U(1) theory, one obtains an IGM dual eq. 3. This fact implies a new view of the nature of the phase transition of pure gauge compact QED. A numerical simulation has been performed [7] observing an apparent first order phase transition, as expected since this model has the same topological structure than pure gauge compact QED.

## 5. FIELD THEORIES OF THE THIRD-RANK

Some years ago, Poliakov pointed out that an abelian gauge theory with fields of the third rank will present, in four dimensions, point-like topological excitations and hence, no phase transitions coming from monopole condensation.

As a first step in the analysis of this question we have studied the four-dimensional abelian Z(2) case [8]. We denote by $n$ a lattice site and by $n_\mu$ a link starting from point $n$ in the direction $\mu$ of an

4-d hypercubical lattice. Starting from this point there are 4 positive links. A simple plaquette defined by directions $\mu$, $\nu$ is denoted by $n_{\mu\nu}$. We assign a gauge variable ($\sigma_i = \pm 1$) to each plaquette. Elementary three-dimensional cubes are composed by six plaquettes ($i \in \partial_{cube}$, i.e. the plaquette belongs to the perimeter of the cube). Three of them are the plaquettes associated to the combinations of the three index $\mu, \nu, \eta$ related to the links starting from a given point $n$ in the positive directions. The remainder three plaquettes are those that "close" the cube. The partition function is defined, then, as:

$$Z = \sum_{\{\sigma_i\}} e^{-\beta E}; \qquad E = \sum_{cube=1}^{DL^D} E(cube); \qquad (4)$$

where

$$E(cube) = 1 - \prod_{i \in \partial_{cube}} \sigma_i; \qquad \sigma_i = \pm 1, \qquad (5)$$

with $\beta > 0$ being the inverse temperature in natural units.

A duality transformation [9] relates this four dimensional gauge model to the four dimensional Ising model, which has a second order phase transition with mean field critical exponents [10].

Our estimates for the critical exponents $\nu$ and $\alpha$ exclude a first order phase transition, and they are consistent with those expected from the 4-d Ising model. However, a plot of the energy histogram just over the phase transition point, shows a clear two-peaks structure. Moreover, a further finite-size analysis of the $P_L^{min}$'s reveals that in the limit $L \to \infty$ the surface tension is compatible with zero, giving place to a continuous transition.

We conclude that the system actually experiments a second order phase transition with mean field critical exponents, as predicted by duality, and that the intriguing double peak structure seems to be due to an added phenomena, that may be understood in terms of the topological excitations of the model, i.e. the monopoles. This image is consistent with the fact than in five dimensions, where the topological excitations become 1-dimensional, this model exhibits a clear first order phase transition.

## 6. CONCLUSIONS

Our main conclusions can be summarized as follows

- Pure gauge compact QED with PBC shows a monopole percolation phenomena coupled to the monopole condensation.

- Pure gauge compact QED with Fixed Boundary Conditions (spherical topology) shows a behaviour compatible with a second order phase transition

- The $\gamma \to \infty$ limit of Non Compact Scalar QED with PBC is related to the Villain form of compact QED and also to the Loop Model for QED, showing a first order phase transition.

- Numerical a analysis of a third rank abelian Z(2) theory shows an intriguing behaviour over the phase transition that may be originated by point-like topological excitations.

The simulations done here used mainly the CRAY C90 at PSC and the CRAY Y/MP of CESCA.